%% file: main.tex
\documentclass[aps,prl,twocolumn,superscriptaddress,longbibliography,floatfix]{revtex4-2}

\usepackage{amsmath,amssymb,bm}
\usepackage{graphicx}
\usepackage{tikz}
\usetikzlibrary{decorations.pathmorphing,decorations.markings,arrows.meta,calc,shapes.misc}
\usepackage{xcolor}
\usepackage[hidelinks]{hyperref}

\newcommand{\rs}{r_s}
\newcommand{\kF}{k_{\mathrm F}}

\newcommand{\Sig}{\Sigma}
\newcommand{\Gamm}{\Gamma}

\newcommand{\Ree}{\mathrm{Re}}

\begin{document}

\title{First-Principles Effective Mass in the Three-Dimensional Uniform Electron Gas}

\author{Pengcheng Hou}
\email{houpc@hfnl.cn}
\thanks{These three authors contributed equally to this paper.}
\affiliation{Hefei National Laboratory, University of Science and Technology of China, Hefei 230088, China}

\author{Daniel Cerkoney}
\thanks{These three authors contributed equally to this paper.}
\affiliation{Department of Physics and Astronomy, Rutgers, The State University of New Jersey, Piscataway, NJ 08854-8019 USA}

\author{Zhiyi Li}
\thanks{These three authors contributed equally to this paper.}
\affiliation{Department of Modern Physics, University of Science and Technology of China, Hefei, Anhui 230026, China}
\affiliation{Hefei National Laboratory, University of Science and Technology of China, Hefei 230088, China}

\author{Tao Wang}
\affiliation{Beijing National Laboratory for Condensed Matter Physics and Institute of Physics, \\Chinese Academy of Sciences, Beijing 100190, China}

\author{Xiansheng Cai}
\affiliation{CAS Key Laboratory of Theoretical Physics, Institute of Theoretical Physics, Chinese Academy of Sciences, Beijing 100190, China}

\author{Lei Wang}
\affiliation{Beijing National Laboratory for Condensed Matter Physics and Institute of Physics, \\Chinese Academy of Sciences, Beijing 100190, China}

\author{Gabriel Kotliar}
\affiliation{Department of Physics and Astronomy, Rutgers, The State University of New Jersey, Piscataway, NJ 08854-8019 USA}

\author{Youjin Deng}
\email{yjdeng@ustc.edu.cn}
\affiliation{Department of Modern Physics, University of Science and Technology of China, Hefei, Anhui 230026, China}
\affiliation{Hefei National Laboratory, University of Science and Technology of China, Hefei 230088, China}

\author{Kun Chen}
\email{chenkun@itp.ac.cn}
\affiliation{CAS Key Laboratory of Theoretical Physics, Institute of Theoretical Physics, Chinese Academy of Sciences, Beijing 100190, China}
\affiliation{Department of Physics and Astronomy, Rutgers, The State University of New Jersey, Piscataway, NJ 08854-8019 USA}
\affiliation{Center for Computational Quantum Physics, Flatiron Institute, 162 5th Avenue, New York, New York 10010}
\date{\today}

\begin{abstract}
    The quasiparticle effective mass $m^*$ of the three-dimensional uniform electron gas is a fundamental Fermi-liquid parameter whose value and density dependence have remained controversial for decades.
    Using renormalized perturbation theory with explicit counterterms, we determine $m^*$ in the metallic regime ($r_s \le 6$) from first principles by two complementary routes---the self-energy and the forward-scattering four-point vertex via the $p$-wave spin-symmetric Landau parameter $F_1^s$---that agree within uncertainties at each density through sixth renormalized order.
    The resulting $m^*/m$ remains close to unity throughout the metallic regime, with a shallow non-monotonic density dependence---a minimum near $r_s\approx 1$ followed by a gentle upturn---reflecting the interplay of exchange and dynamical screening in the self-energy, and disfavoring strong monotonic suppression.
    This near-unity behavior is traced to the structural dominance of long-range charge fluctuations, whose isotropy in angular-momentum space suppresses $F_1^s$ that controls $m^*/m$.
\end{abstract}

\maketitle

\textbf{Introduction.---}
The three-dimensional uniform electron gas (3D UEG) is the paradigmatic correlated Fermi system.
As the simplest setting in which long-ranged Coulomb interactions act in a translationally invariant quantum many-body ground state, it provides a clean reference for correlation physics in simple metals.
Accordingly, it underpins exchange--correlation functionals, calibrates many-body approximations used in electronic-structure calculations, and serves as a stringent test of controlled numerical methods~\cite{CeperleyAlder1980,loos2016uniform,dornheim2018uniform}.

In the metallic regime, the UEG is expected to be a Landau Fermi liquid, whose low-energy properties are encoded in a small set of parameters: the quasiparticle residue $Z$, the effective mass ratio $m^*/m$ (equivalently the renormalized Fermi velocity $v_F^*=\kF/m^*$, with $\kF$ the Fermi wavevector), and the Landau interaction function, parametrized by its angular-momentum harmonics $F_l^{s/a}$~\cite{pines2018theory,GiulianiVignale2005,baym2008landau}.
Among these quantities, the effective mass is especially important because it governs the low-energy quasiparticle dynamics near the Fermi surface and enters directly into the electronic specific heat, spin susceptibility, and other thermodynamic properties~\cite{nozieres2018theory,GiulianiVignale2005}.
It is also particularly demanding to determine reliably: it must be inferred from delicate low-energy structure near the Fermi surface and is therefore sensitive to how screening and vertex corrections are controlled~\cite{simion2008many,eich_effective_2017}.

\begin{figure}[t]
    \centering
    \includegraphics[width=0.88\columnwidth]{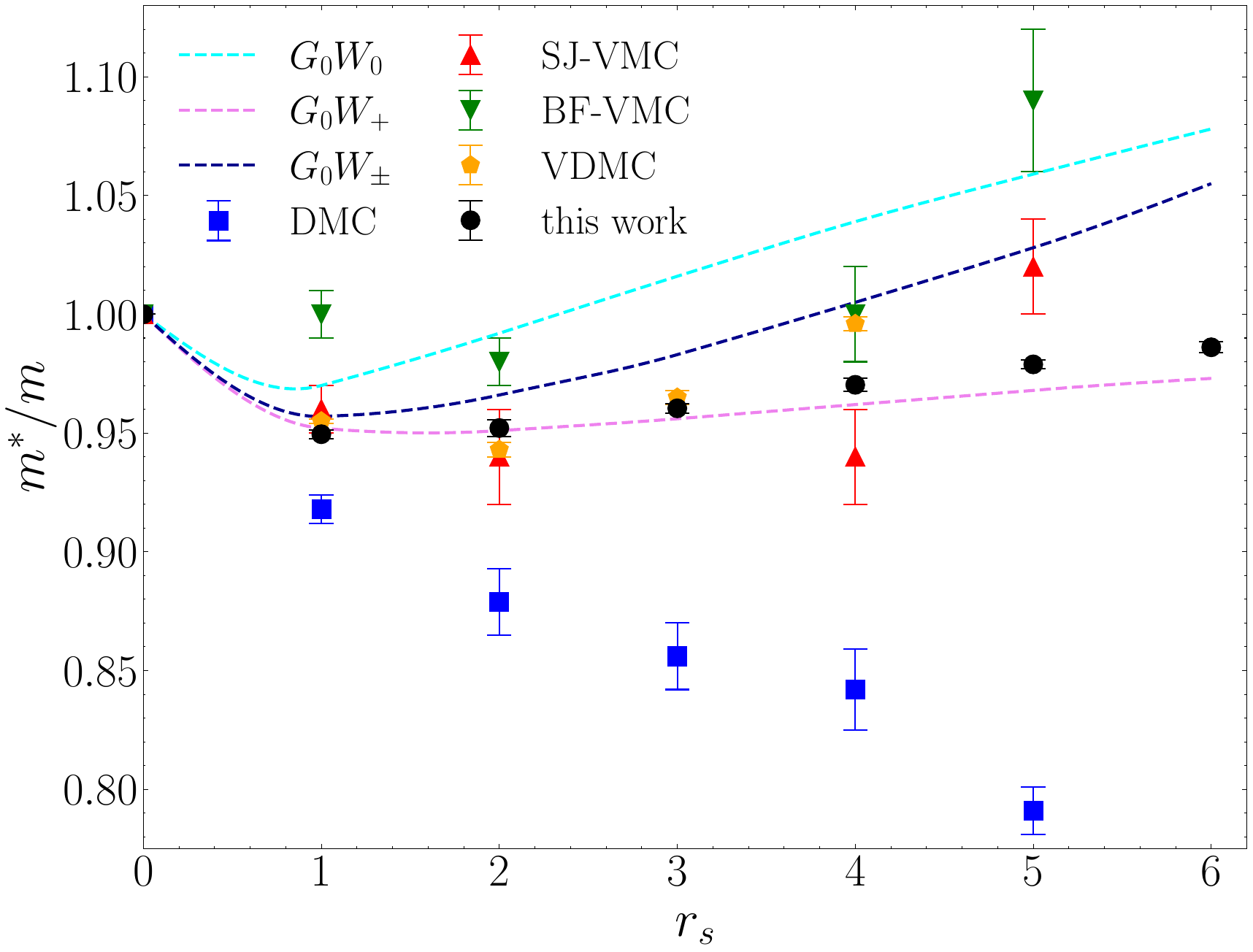}
    \caption{\label{fig:meff}
    Quasiparticle effective mass ratio $m^*/m$ of the uniform electron gas versus $\rs$. Black circles show the inverse-variance weighted means of our self-energy and vertex determinations, with error bars conservatively set by the larger route uncertainty. The route-resolved values are listed in Table~\ref{tab:compare}.
    Other symbols show representative literature estimates:
    variational diagrammatic Monte Carlo (VDMC)~\cite{haule2022single}, diffusion Monte Carlo (DMC)~\cite{azadi2021},
    and variational Monte Carlo (VMC) using Slater--Jastrow (SJ) and backflow (BF) trial wave functions~\cite{holzmann2023static}.
    Dashed curves illustrate standard diagrammatic approximations: $G_0W_0$ and two vertex-improved variants ($G_0W_+$ and $G_0W_{\pm}$)~\cite{simion2008many}.
    The result remains near unity for $\rs\le 6$ and disfavors a strong monotonic suppression over this density range.
    }
\end{figure}

Consequently, whether interactions enhance or suppress the quasiparticle mass at metallic densities remains a long-standing question, with different many-body approaches yielding conflicting density trends.
Low-order diagrammatic approaches based on random-phase-approximation (RPA) screening [e.g., $G_0W_0$] typically yield a modest mass enhancement at metallic densities \cite{QuinnFerrell1958,rice1965effects,hedin1969solid,Takada1991,HolmVonBarth1998}, while incorporating vertex corrections via local-field factors~\cite{KukkonenOverhauser1979,ng1986effective1,yasuhara1992effective,PhysRevB.53.7352,simion2008many} can substantially modify both the magnitude and the curvature of $m^*(\rs)$~\cite{simion2008many} ($\rs$ is the Wigner--Seitz density parameter).
On the numerical side, different quantum Monte Carlo (QMC) strategies~\cite{foulkes2001quantum} have led to qualitatively different conclusions: diffusion Monte Carlo (DMC) estimates extracted from finite-size excitation energies indicate a pronounced monotonic suppression of $m^*/m$ with increasing $\rs$~\cite{azadi2021}, whereas the variational diagrammatic Monte Carlo (VDMC) self-energy calculation of Ref.~\cite{haule2022single} supports a near-unity, non-monotonic density dependence. Variational Monte Carlo (VMC) estimates obtained from the static self-energy are also broadly near unity but carry comparatively larger uncertainties~\cite{holzmann2023static} and therefore do not resolve the incompatible DMC and VDMC trends (Fig.~\ref{fig:meff}).
Analogous discrepancies persist in the two-dimensional electron gas, where different QMC protocols and variational approaches yield conflicting trends for $m^*(\rs)$~\cite{drummond2009quantum,drummond2013diffusion,xie2023m,azadi2025qmc2d}, underscoring that the challenge of reliably extracting the effective mass extends across dimensions.
The coexistence of these incompatible trends motivates a controlled determination that can separate physical behavior from methodological systematics.

Beyond the spread of reported values, there is a more structural facet:
in an exact Galilean-invariant Fermi liquid, different routes to the effective mass are not independent, but are tied together by conservation laws (Ward identities).
In particular, the effective mass obtained from the low-energy quasiparticle dispersion (a single-particle characterization) must coincide with the mass inferred from the forward-scattering four-point vertex via the $p$-wave ($l{=}1$) spin-symmetric Landau parameter $F_1^s$ (a two-particle characterization).
This linkage is guaranteed when self-energy and vertex are treated consistently~\cite{Baym1961,Baym1962},
as emphasized for jellium in vertex-corrected $GW$-type studies~\cite{HindgrenAlmbladh1997}.
Conversely, approximate schemes can violate the linkage if the vertex content is incomplete or inconsistent with the self-energy, obscuring whether different calculations are targeting the same low-energy parameter.
A controlled determination therefore calls for a framework in which diagrammatic content is systematically controlled and the self-energy and vertex routes can be evaluated on the same footing.

In this Letter, we provide such a determination.
We compute $m^*/m$ for the paramagnetic 3D UEG in the metallic regime ($\rs\le 6$) using a renormalized perturbation theory defined directly from a renormalized action with explicit counterterms,
where all diagrams at a given renormalized order are included and evaluated from first principles.
Within this same expansion, we extract $m^*$ by two complementary routes: (i) from the quasiparticle dispersion via the self-energy and (ii) from the Landau parameter $F_1^s$ via the forward-scattering four-point vertex.
The two routes agree within uncertainties at each density, with their combined estimates shown in Fig.~\ref{fig:meff} and the route-resolved values listed in Table~\ref{tab:compare}.
Our results show that $m^*/m$ remains close to unity across the metallic regime, with a shallow non-monotonic density dependence including a minimum near $\rs\approx 1$.
We trace this near-unity behavior to the structural dominance of long-range charge fluctuations, which lock the frequency and momentum renormalizations of the self-energy together and suppress the anisotropic ($p$-wave) Landau interaction that controls $m^*/m$.

\begin{table*}[t]
    \caption{
    Representative values of the effective-mass ratio $m^*/m$ in the uniform electron gas at selected $\rs$ from different approaches.
    ``This work'' lists the two complementary determinations:
    from the self-energy ($\Sigma$ route) and from the forward-scattering four-point vertex ($\Gamma_4$ route via $F_1^s$).
    Variational diagrammatic Monte Carlo (VDMC) denotes the Hedin-type self-energy result of Ref.~\cite{haule2022single}.
    BF-VMC and SJ-VMC denote the variational Monte Carlo static self-energy analysis using backflow (BF) and Slater-Jastrow (SJ) trial wave functions~\cite{holzmann2023static}.
    Low-order diagrammatic entries show $G_0W_0$ (RPA screening) and two commonly used vertex-improved variants ($G_0W_{+}$ and $G_0W_{\pm}$)~\cite{simion2008many}. DMC values are digitized from Fig.~4 of Ref.~\cite{azadi2021}
    and are quoted at the precision of the original figure.
    }
    \begin{ruledtabular}
        \scriptsize
        \begin{tabular}{c c c c c c c c c c}
            $\rs$                            &
            This work ($\Sigma$)             &
            This work ($\Gamma_4$)           &
            VDMC~\cite{haule2022single}      &
            DMC~\cite{azadi2021}             &
            BF-VMC~\cite{holzmann2023static} &
            SJ-VMC~\cite{holzmann2023static} &
            $G_0W_0$~\cite{simion2008many}   &
            $G_0W_+$~\cite{simion2008many}   &
            $G_0W_{\pm}$~\cite{simion2008many}                                                                                            \\
            \hline
            1                                & 0.9495(43) & 0.9498(20) & 0.955(1) & 0.918(6)  & 1.00(1) & 0.96(1) & 0.970 & 0.952 & 0.957 \\
            2                                & 0.9521(35) & 0.9521(39) & 0.943(3) & 0.879(14) & 0.98(1) & 0.94(2) & 0.992 & 0.951 & 0.966 \\
            3                                & 0.9604(20) & 0.9607(30) & 0.965(3) & 0.856(14) & --      & --      & 1.016 & 0.956 & 0.983 \\
            4                                & 0.9703(27) & 0.9701(31) & 0.996(3) & 0.842(17) & 1.00(2) & 0.94(2) & 1.039 & 0.962 & 1.005 \\
            5                                & 0.9789(18) & 0.9784(23) & --       & 0.791(10) & 1.09(3) & 1.02(2) & 1.059 & 0.968 & 1.028 \\
            6                                & 0.9861(23) & 0.9853(18) & --       & --        & --      & --      & 1.078 & 0.973 & 1.055 \\
        \end{tabular}
    \end{ruledtabular}
    \label{tab:compare}
\end{table*}

\textbf{Two Fermi-liquid routes to the effective mass.---}
As discussed above, two distinct Fermi-liquid relations provide complementary determinations of $m^*$, and their mutual agreement directly tests the internal consistency of the diagrammatic content~\cite{pines2018theory,Baym1961}.

The first route is single-particle: one extracts the renormalized quasiparticle dispersion
in the vicinity of the Fermi surface from the self-energy and determines its slope at $k=\kF$.
The result can be expressed in terms of the self-energy on the imaginary-frequency axis,
\begin{equation}
    \label{eq:mstarSigma}
    \frac{m^*}{m}
    =
    Z^{-1}\Bigl[1+\frac{m}{\kF}\,\partial_k \Ree \Sig(k,i\omega)\big|_{\kF,0}\Bigr]^{-1} \,,
\end{equation}
where $Z=\left[1-\partial_\omega \mathrm{Im} \Sig(k,i\omega)\big|_{\kF,0}\right]^{-1}$. We denote the quantity in square brackets by $D$, the momentum-renormalization factor, so that $m/m^*=Z D$. Both derivatives denote the $T\to0$ limits at $k=\kF$ and $\omega=0$.

The second route is two-particle.
We start from the proper four-point vertex in the forward-scattering (``$\omega$'') limit and define the Landau quasiparticle interaction function on the Fermi surface as
\begin{equation}
    \begin{aligned}
        \label{eq:fGamma}
        f_{\sigma\sigma'}(\theta) & =Z^2\,\Gamm^{\omega}_{\sigma\sigma'}(\kF,\kF;\theta)                                                                                                                                                           \\
                                  & =Z^2 \lim _{\mathbf{q} \rightarrow 0}\left[\Gamma_{\sigma \sigma^{\prime}}\left(\mathbf{k}_{\mathbf{1}} \omega_{-1}, \mathbf{k}_2 \omega_0 ; \mathbf{k}_{\mathbf{1}}+\mathbf{q} \omega_0\right)-V_q\right] \,,
    \end{aligned}
\end{equation}
where $\theta$ is the angle between the incoming momentum $\mathbf k_1$ and $\mathbf k_2$ on the Fermi surface, $\omega_0 = i\pi T$ and $\omega_{-1} = -i\pi T$ are the lowest fermionic Matsubara frequencies, and $V_q=4\pi e^2/q^2$ is the Coulomb interaction.
Writing $f_{\rm s}(\theta)\equiv [f_{\uparrow\uparrow}(\theta)+f_{\uparrow\downarrow}(\theta)]/2$, let $N_F=mk_F/\pi^2$ be the total bare density of states and $N_F^*=(m^*/m)N_F$ the quasiparticle density of states. We use the convention $F_1^s=N_F^*\langle f_{\rm s}(\theta)\cos\theta\rangle$, for which Galilean invariance gives $m^*/m=1+F_1^s$~\cite{coleman,pines2018theory}. (With the conventional Legendre normalization $F_{1,\mathrm{std}}^s=3F_1^s$, this is the familiar $m^*/m=1+F_{1,\mathrm{std}}^s/3$.) Eliminating $N_F^*$ gives the vertex-route mass as
\begin{equation}
    \label{eq:mstarGammaCos}
    \frac{m^*}{m}=\frac{1}{1-N_F\left\langle f_{\rm s}(\theta)\cos\theta\right\rangle}
        =\frac{1}{1-N_F Z^2 \left\langle \Gamma^\omega_{4,\rm s}(\theta)\cos\theta \right\rangle} \, .
\end{equation}
Here $\Gamma^\omega_{4,\rm s}(\theta)\equiv [\Gamma^\omega_{\uparrow\uparrow}(\theta)+\Gamma^\omega_{\uparrow\downarrow}(\theta)]/2$, and $\langle\cdots\rangle$ is the normalized Fermi-surface angular average.

The Ward-identity equivalence of the two routes holds exactly only when self-energy and vertex are treated on a fully consistent footing.
In the present work, both routes are evaluated within the same renormalized perturbation theory and computed with the same diagrammatic machinery, so their mutual agreement constitutes a parameter-free internal consistency check.

\textbf{Method.---}
We evaluate both the self-energy and four-point-vertex routes within the same renormalized perturbation theory defined directly from a renormalized action with explicit counterterms.
We expand around a theory with a renormalized chemical potential $\mu_R$ and a statically screened Yukawa interaction $V_q(\lambda_R)=4\pi e^2/(q^2+\lambda_R)$, treating deviations from the physical Coulomb problem as counterterms.
The chemical-potential counterterm $\delta\mu \equiv \mu-\mu_R$ is expanded order by order in a bookkeeping parameter $\xi$ (renormalized order),
$\delta\mu(\xi)=\sum_{p\ge 1}\xi^p\,\delta\mu^{(p)}$,
and its coefficients are fixed by requiring that the interacting Fermi surface remain unshifted at each order.
In addition, we reexpand the bare Coulomb interaction as $\frac{4 \pi e^2}{q^2}=V_q(\lambda_R) \sum_{n=0}^{\infty} \left(\xi\frac{\lambda_R}{4 \pi e^2} V_q(\lambda_R)\right)^n$,
so that $\lambda_R$ acts only as an auxiliary reorganization parameter and the physical Coulomb interaction is recovered at $\xi=1$ when the reexpanded series is summed.
At each density, we identify a finite-order stability window using the order-by-order behavior and the variation of $m^*/m$ with $\lambda_R$ in both routes. The resulting windows satisfy $\sqrt{\lambda_R}=O(q_{\rm TF})$, where $q_{\rm TF}^2=4\pi e^2N_F\simeq2.44/(\rs a_B^2)$ and $a_B$ is the Bohr radius. This correspondence provides a natural physical scale for the magnitude and density dependence of the finite-order stability range without implying a unique optimum. Representative scans and the stability windows used at all densities are provided in the Supplemental Material (SM)~\cite{SM}.

With this reorganization, the expansion at each order comprises renormalized Feynman diagrams built from $V_q(\lambda_R)$ and the reference Green's function $G_0(\mu_R)$, together with all counterterm insertions ($\delta\mu$ on fermion lines, $\frac{\lambda_R}{4 \pi e^2}$ on interaction lines) generated by Taylor-expanding lower-order diagrams, as illustrated schematically in Fig.~\ref{fig:diagrams}.

\begin{figure}
    \centering
    \input{fig_diagrams.tex}
    \caption{\label{fig:diagrams}
        Schematic structure of the renormalized diagrammatic expansion, organized by powers of the bookkeeping parameter $\xi$.
        (a)~Self-energy $\Sig$: the first-order Fock diagram and representative second-order contributions, including skeleton diagrams (two Yukawa lines) and counterterm insertions on the first-order graph.
        (b)~Four-point vertex $\Gamm_4$: the zeroth-order (tree-level) Yukawa exchange, representative first-order skeleton diagrams (particle-hole bubble and particle-particle ladder), and counterterm insertions.
        Solid lines with arrows denote the reference Green's function $G_0(\mu_R)$; wavy lines denote the Yukawa interaction $V_q(\lambda_R)$; crosses ($\times$) and filled squares ($\blacksquare$) mark the chemical-potential counterterm $\delta\mu$ and the interaction counterterm $\lambda_R/(4\pi e^2)$, respectively.
        At each renormalized order, all skeleton topologies and their counterterm insertions are generated automatically by Taylor-mode AD and included without truncation.
    }
\end{figure}

All counterterm bookkeeping and renormalized-series generation are implemented automatically by Taylor-mode automatic differentiation (AD)~\cite{Taylor1,Taylor2,Taylor3} on diagrammatic computational graphs. Diagram generation and numerical kernel construction are carried out within our \texttt{FeynmanDiagram.jl} framework~\cite{HouTechStack2025}, which translates renormalized diagram classes into optimized representations for Taylor-mode AD and Monte Carlo evaluation via \texttt{MCIntegration.jl}~\cite{MCIntegration}. At each density $r_s$ we compute the renormalized series through sixth order.
Both $\Sigma$ and $\Gamma_4$ are generated directly from the bare Coulomb action within the same explicit Taylor-reexpanded series, enabling an order-by-order consistency test between the single- and two-particle routes. No screened interaction supplied by a separate calculation enters the present expansion as an intermediate input. The VDMC calculation of Ref.~\cite{haule2022single}, by comparison, evaluates a Hedin-type self-energy using the screened interaction obtained from the separate response-function calculation of Ref.~\cite{chen2019combined}.

The self-energy derivatives entering Eq.~\eqref{eq:mstarSigma} are extracted from Matsubara-frequency data computed at several low temperatures and extrapolated to $T=0$. The forward-scattering vertex entering Eq.~\eqref{eq:mstarGammaCos} is obtained by evaluating $\Gamma_{\sigma\sigma'}$ at the two lowest fermionic Matsubara frequencies and taking $q=0$.
At each density and renormalized order, the quoted uncertainties combine in quadrature Monte Carlo statistical errors, estimates of the truncation error from the maximum variation across the last three renormalized orders ($N=4,5,6$), the systematic uncertainty of the $T\to0$ extrapolation, and the residual sensitivity to the auxiliary screening parameter $\lambda_R$.
The Supplemental Material provides these details.

\textbf{Results.---}
As shown in Fig.~\ref{fig:meff}, $m^*/m$ remains close to unity for $r_s \le 6$, with a shallow minimum near $\rs\approx1$ followed by a gentle increase toward larger $\rs$.
The two complementary determinations are consistent within uncertainties at each $\rs$. Figure~\ref{fig:meff} shows their combined estimates, while Table~\ref{tab:compare} lists the route-resolved values.

Figure~\ref{fig:meff} and Table~\ref{tab:compare} place our results in the context of representative diagrammatic approximations and QMC-based extractions.
At metallic densities, low-order $G_0W_0$ and commonly used vertex-improved $G_0W$ variants yield only modest renormalization of $m^*/m$, whereas published QMC estimates span from near-unity values to a pronounced monotonic suppression with increasing $\rs$.
The VDMC results of Ref.~\cite{haule2022single} are broadly consistent with ours for $\rs\le3$. At $\rs=4$, however, both our self-energy and vertex routes yield $m^*/m=0.970(3)$, below the VDMC value $0.996(3)$. The agreement of the two complementary routes shows that the lower value is not specific to either Fermi-liquid extraction.
The order stabilization is illustrated representatively in Fig.~\ref{fig:convComb}. The residual $\lambda_R$ sensitivity is documented in the Supplemental Material, where variations across the identified stability windows remain below $1\%$ and are included in the quoted uncertainties.

Figure~\ref{fig:convComb} provides a representative order-by-order view at $\rs=5$, a challenging metallic density.
In the self-energy route [panel~(a)], $Z$ and $D$ each continue to evolve through sixth order without clear convergence, while their product $m/m^*=Z D$ reaches a stable value by fourth or fifth order, showing that their remaining order-dependent changes largely compensate in the mass combination.
In the vertex route [panel~(b)], the raw two-particle ingredients likewise continue to evolve with order, while the derived low-energy mass ratio obtained from Eq.~(\ref{eq:mstarGammaCos}) approaches a stable plateau appreciably earlier.
Thus, although intermediate quantities remain sensitive to short-range correlation physics, the effective-mass ratio itself is already well controlled at the orders we reach.
Order-by-order convergence data for both routes are provided in the Supplemental Material.

\begin{figure}[tb]
    \centering
    \includegraphics[width=0.98\columnwidth]{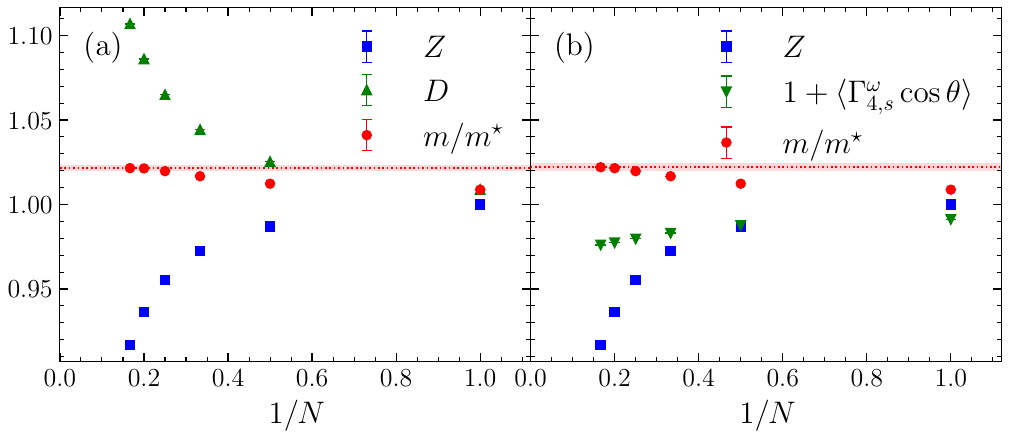}
    \caption{\label{fig:convComb}
    Renormalized-order stabilization of $m^*/m$ at $\rs=5$ ($\lambda_R=1.375\,a_B^{-2}$), a representative challenging metallic density.
    (a)~Self-energy route: the quasiparticle residue $Z$ and momentum-renormalization factor $D$ each evolve substantially with order without reaching clear convergence at sixth order, while their product $m/m^*=Z\cdot D$ [Eq.~(\ref{eq:mstarSigma})] stabilizes much earlier.
    (b)~Vertex route: $Z$ and the shifted forward-scattering quantity $1+\langle\Gamm_{4,\rm s}^\omega(\theta)\cos\theta\rangle$ (shown for visualization) continue to renormalize, whereas the derived mass ratio $m/m^*$ evaluated from the full combination in Eq.~(\ref{eq:mstarGammaCos}) reaches a stable plateau appreciably earlier.
    The shaded band indicates the final estimate and its uncertainty.
    }
\end{figure}

The near-unity mass ratio throughout the metallic regime reflects a structural property of the electron-gas self-energy: the dominant many-body effects---charge screening, exchange--correlation hole formation, and plasmon exchange---are mediated by long-range charge fluctuations peaked at small momentum transfer $q\ll\kF$ and are therefore largely isotropic in angular-momentum space.
In the self-energy identity $m^*/m=(Z\cdot D)^{-1}$, this forward-scattering dominance is directly visible: at small $q$, the internal propagator satisfies $(m/\kF)\,\partial_k G_0\approx\partial_\omega G_0$, locking the frequency and momentum renormalizations of $\Sigma$ so that $Z^{-1}\approx D$ up to corrections of order $\langle q/\kF\rangle_W$, as already manifest in Hedin's $G_0W_0$ analysis~\cite{hedin1965new}; although $Z$ and $D$ each deviate substantially from unity (Fig.~\ref{fig:convComb}a), their product stays close to~1, i.e., the on-shell self-energy $\Sigma(k,\varepsilon_k)$ has negligible slope at $\kF$.
In the Landau framework, the same isotropy implies that the dominant correlations feed the $s$-wave channel rather than the $p$-wave parameter $F_1^s$ that controls $m^*/m$~\cite{li2025twoelectron}, structurally disfavoring large mass renormalization~\cite{simion2008many}.
Beyond $G_0W_0$, vertex insertions modify the integrands but preserve the long-range screening that enforces forward-scattering dominance; our sixth-order data confirm that the $Z$--$D$ near-cancellation persists, providing direct evidence that this structural protection extends to high order.

The shallow minimum near $\rs\approx 1$ and the gentle upturn at larger $\rs$ reflect the evolving competition between exchange and correlation contributions to the self-energy.
At high density, exchange steepens the quasiparticle dispersion (reducing $m^*/m$ below unity), while at lower density dynamical screening grows and its contribution to the self-energy momentum derivative opposes exchange, gradually pushing $m^*/m$ back toward unity over the range studied.
Our controlled high-order results place this competition quantitatively and confirm that the non-monotonic behavior is a genuine feature of the 3D UEG, not an artifact of low-order truncation.
The strong monotonic suppression reported in the DMC extraction of Ref.~\cite{azadi2021} differs qualitatively from this pattern as well as from the near-unity VMC and VDMC values~\cite{holzmann2023static,haule2022single}.
In the DMC study of Ref.~\cite{azadi2021}, $m^*/m$ is inferred from the Fermi-surface slope of a global Pad\'e fit to finite-cell $(N{\pm}1)$-electron addition and removal energies, followed by extrapolation to the thermodynamic limit. The central methodological question is whether the Fermi-surface slope obtained from this broad-window finite-cell fit converges to that of the thermodynamic-limit quasiparticle dispersion. Because the fit is constrained by excitations away from $\kF$, this correspondence has not been demonstrated at the precision relevant to the present discrepancy~\cite{holzmann2023static}, and a companion two-dimensional DMC study likewise identifies this correspondence as an open question~\cite{azadi2025qmc2d}. The SM~\cite{SM} examines this extraction together with its finite-size, trial-state, and algorithmic sensitivities.

\textbf{Conclusion.---} We have presented a controlled first-principles determination of the quasiparticle effective mass of the 3D paramagnetic UEG in the metallic regime ($\rs\le 6$), computed through sixth order in a renormalized perturbation theory with Taylor-mode AD on diagrammatic computational graphs.
The two complementary Fermi-liquid routes---self-energy and forward-scattering vertex---agree within uncertainties at each density; the resulting $m^*/m$ lies in the near-unity range with a shallow non-monotonic density dependence, disfavoring a strong monotonic suppression.
This agreement provides a stringent internal check on the diagrammatic construction and renormalization procedure.
More broadly, the present diagrammatic framework enables controlled access to vertex-level Fermi-liquid parameters in Coulomb systems~\cite{li2025twoelectron} and can be extended to other observables and correlated-electron reference problems.

A direct experimental test of the ideal 3D UEG remains difficult: real metals retain the underlying Coulomb interaction but introduce band-structure, electron--phonon, and disorder effects, whereas current ultracold atomic Fermi-gas platforms offer cleaner control but generally realize short-ranged interactions rather than the $1/r$ interaction of the UEG. Low-temperature calorimetry nevertheless offers an indirect thermodynamic constraint. For K and Rb ($\rs\simeq4.9$ and $5.2$), combining the thermal mass $m_{\rm th}$ inferred from the measured specific heat with pseudopotential estimates of the band mass $m_{\rm b}$ and electron--phonon mass-enhancement parameter $\lambda_{\rm ep}$ yields $m_{\rm th}/[m_{\rm b}(1+\lambda_{\rm ep})]\simeq1.0$ as the inferred electron--electron contribution to the mass ratio in each metal~\cite{LienPhillips1964,Grimvall1975,Grimvall1981}. Within the same decomposition, assigning a mass ratio near $0.8$ to the electron--electron contribution would require a substantially larger electron--phonon enhancement than inferred for these weak-coupling alkali metals. Although model dependent, this thermodynamic comparison favors a near-unity electron--electron contribution to the mass ratio over the ${\sim}20\%$ suppression found in DMC~\cite{azadi2021}. More precise calorimetry and improved material-specific estimates of $m_{\rm b}$ and $\lambda_{\rm ep}$ could sharpen this indirect test.

\begin{acknowledgments}
    We thank S. Das Sarma for useful discussions.
    P.H., Z.L., and Y.D. were supported by the National Natural Science Foundation of China (under Grant No. 12275263) and the Quantum Science and Technology-National Science and Technology Major Project (under Grant No. 2021ZD0301900).
    D.C. was supported by the US Department of Energy, Office of Basic Energy Sciences as part of the Computation Material Science Program.
    L.W. is supported by the National Natural Science Foundation of China under Grants No. T2225018, No. 92270107, and No. 12188101, No. T2121001, and the Strategic Priority Research Program of Chinese Academy of Sciences under Grants No. XDB0500000 and No. XDB30000000.
    K.C. was supported by the National Key Research and Development Program of China, Grant No. 2024YFA1408604, the National Natural Science Foundation of China under Grants No. 12474245 and No. 12447103, and the GHfund A(202407010637).

\end{acknowledgments}

\section{Data Availability}
The data that support the findings of this article are openly available~\cite{Hou_Data_for_First-Principles_2026}.

\nocite{ChubukovMaslov2003,ChubukovMaslovMillis2006,holzmann2011finite}

\bibliographystyle{apsrev4-2}
\bibliography{references}

\end{document}

\title{Supplemental Material for ``First-Principles Effective Mass in the Three-Dimensional Uniform Electron Gas''}
\author{Pengcheng Hou}
\email{houpc@hfnl.cn}
\thanks{These three authors contributed equally to this paper.}
\affiliation{Hefei National Laboratory, University of Science and Technology of China, Hefei 230088, China}

\author{Daniel Cerkoney}
\thanks{These three authors contributed equally to this paper.}
\affiliation{Department of Physics and Astronomy, Rutgers, The State University of New Jersey, Piscataway, NJ 08854-8019 USA}

\author{Zhiyi Li}
\thanks{These three authors contributed equally to this paper.}
\affiliation{Department of Modern Physics, University of Science and Technology of China, Hefei, Anhui 230026, China}
\affiliation{Hefei National Laboratory, University of Science and Technology of China, Hefei 230088, China}

\author{Tao Wang}
\affiliation{Beijing National Laboratory for Condensed Matter Physics and Institute of Physics, \\Chinese Academy of Sciences, Beijing 100190, China}

\author{Xiansheng Cai}
\affiliation{CAS Key Laboratory of Theoretical Physics, Institute of Theoretical Physics, Chinese Academy of Sciences, Beijing 100190, China}

\author{Lei Wang}
\affiliation{Beijing National Laboratory for Condensed Matter Physics and Institute of Physics, \\Chinese Academy of Sciences, Beijing 100190, China}

\author{Gabriel Kotliar}
\affiliation{Department of Physics and Astronomy, Rutgers, The State University of New Jersey, Piscataway, NJ 08854-8019 USA}

\author{Youjin Deng}
\email{yjdeng@ustc.edu.cn}
\affiliation{Department of Modern Physics, University of Science and Technology of China, Hefei, Anhui 230026, China}
\affiliation{Hefei National Laboratory, University of Science and Technology of China, Hefei 230088, China}

\author{Kun Chen}
\email{chenkun@itp.ac.cn}
\affiliation{CAS Key Laboratory of Theoretical Physics, Institute of Theoretical Physics, Chinese Academy of Sciences, Beijing 100190, China}
\affiliation{Department of Physics and Astronomy, Rutgers, The State University of New Jersey, Piscataway, NJ 08854-8019 USA}
\affiliation{Center for Computational Quantum Physics, Flatiron Institute, 162 5th Avenue, New York, New York 10010}
\date{\today}

\maketitle

\setcounter{equation}{0}
\setcounter{figure}{0}
\setcounter{table}{0}
\renewcommand{\theequation}{S\arabic{equation}}
\renewcommand{\thefigure}{S\arabic{figure}}
\renewcommand{\thetable}{S\arabic{table}}

\section{Overview}

This Supplemental Material provides technical details supporting the effective-mass determinations reported in the main text.
We document the order-by-order convergence behavior of the renormalized diagrammatic series, the low-temperature extrapolation protocol for extracting zero-temperature Fermi-liquid parameters, the robustness of the results with respect to the auxiliary screening parameter $\lambda_R$, and the uncertainty budget.
All results presented here are for representative densities; the convergence patterns and systematic-error estimates are qualitatively similar at all densities studied ($\rs = 1, 2, 3, 4, 5, 6$).
In addition, the final section provides a quantitative analysis of the systematic sensitivities of the finite-cell excitation-energy extraction of $m^*$ employed in the diffusion Monte Carlo (DMC) study of Ref.~\cite{azadi2021}.

\section{Order-by-order convergence}

\begin{figure}[t]
    \centering
    \includegraphics[width=0.95\columnwidth]{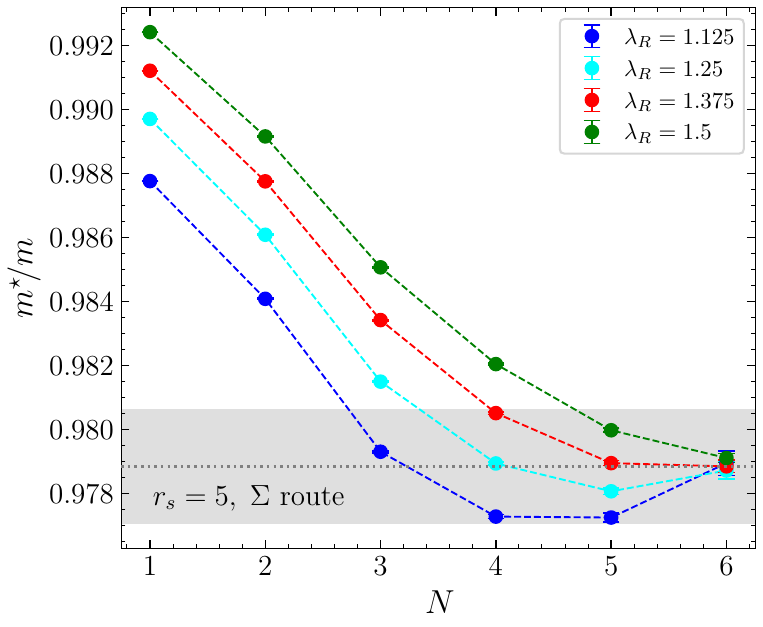}
    \caption{\label{fig:conv_sigma}
        Order-by-order convergence of $m^*/m$ from the self-energy route at $\rs=5$ for several values of the auxiliary screening parameter $\lambda_R$.
        The shaded band indicates the final estimate and its uncertainty.
        At sixth order, the results for the displayed values of $\lambda_R$ agree within this band.
    }
\end{figure}

\begin{figure}[t]
    \centering
    \includegraphics[width=0.95\columnwidth]{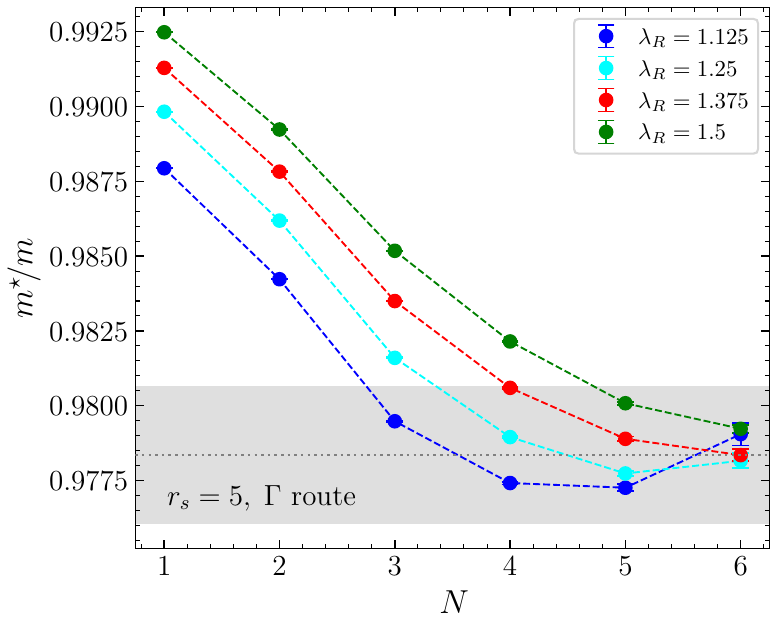}
    \caption{\label{fig:conv_vertex}
        Order-by-order convergence of $m^*/m$ from the vertex route at $\rs=5$ for several values of $\lambda_R$.
        The shaded band indicates the final estimate and its uncertainty. At sixth order, the results for the displayed values of $\lambda_R$ agree within this band.
    }
\end{figure}

Figures~\ref{fig:conv_sigma} and~\ref{fig:conv_vertex} demonstrate the convergence of the effective-mass ratio $m^*/m$ with renormalized perturbation order at $\rs=5$, a representative challenging metallic density, for the self-energy and vertex routes respectively.
Both routes are computed at $T/\TF = 1/40$ with the auxiliary screening parameter $\lambda_R = 1.375$.
Here and throughout the Supplemental Material, numerical values of $\lambda_R$ are quoted in units of $a_B^{-2}$.

As discussed in the main text, the key observable $m^*/m$ stabilizes appreciably earlier than the intermediate quantities from which it is constructed.
In the self-energy route, the quasiparticle residue $Z$ and the momentum-renormalization factor $D$ each evolve substantially through sixth order, yet their order-dependent changes largely compensate in the product $m/m^*=Z\cdot D$, which reaches a stable plateau by fourth or fifth order.
Similarly, in the vertex route, the raw forward-scattering vertex components continue to renormalize, but the combination entering the Landau parameter $F_1^s$ stabilizes early.
This early stabilization indicates correlated renormalization among the ingredients of the physical observable. The agreement between the two routes provides the separate Ward-identity consistency test.

The truncation error at sixth order is estimated from the variation across the last three perturbation orders ($N=4, 5, 6$).
Relying solely on the difference between the final two orders can be misleading for certain choices of $\lambda_R$, where the series may exhibit fortuitously small corrections at a particular order.
A three-point analysis provides a more robust and conservative estimate of the systematic uncertainty associated with series truncation.
We apply this procedure to data at several well-behaved $\lambda_R$ values across the stability plateau around $1.375$ and take the maximum variation as the truncation-error contribution to the final uncertainty budget.

The convergence behavior at the other densities studied ($\rs = 1, 2, 3, 4, 6$) is qualitatively similar: all densities reach stable plateaus by sixth order within comparable relative uncertainties, and the two routes agree at each density.

\section{Low-temperature extrapolation}\label{sec:lowT}

\begin{figure}[t]
    \centering
    \includegraphics[width=0.95\columnwidth]{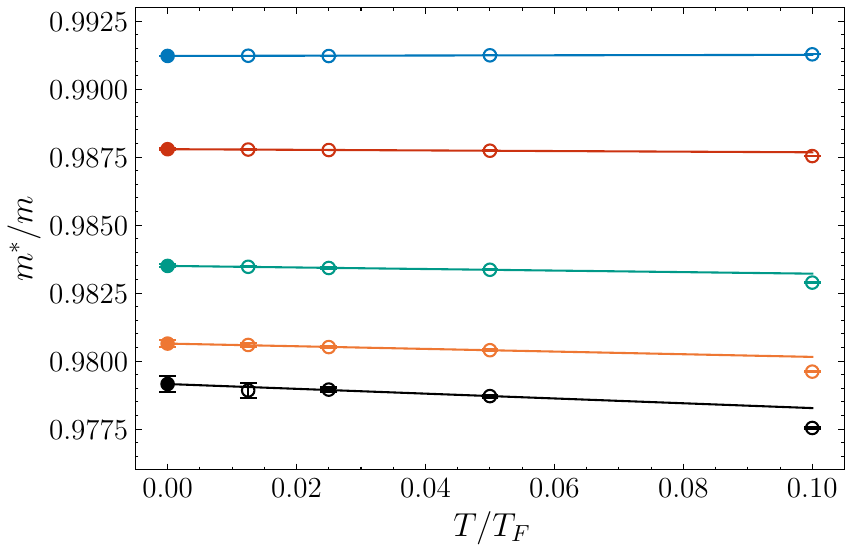}
    \caption{\label{fig:Tscale_sigma}
        Temperature dependence of $m^*/m$ from the self-energy route at $\rs=5$ ($\lambda_R=1.375$) for several renormalized orders.
        The linear-$T$ behavior at low temperature supports the extrapolation protocol used to reach $T=0$.
        The small slopes demonstrate that thermal corrections remain modest throughout the studied temperature range.
    }
\end{figure}

\begin{figure}[t]
    \centering
    \includegraphics[width=0.95\columnwidth]{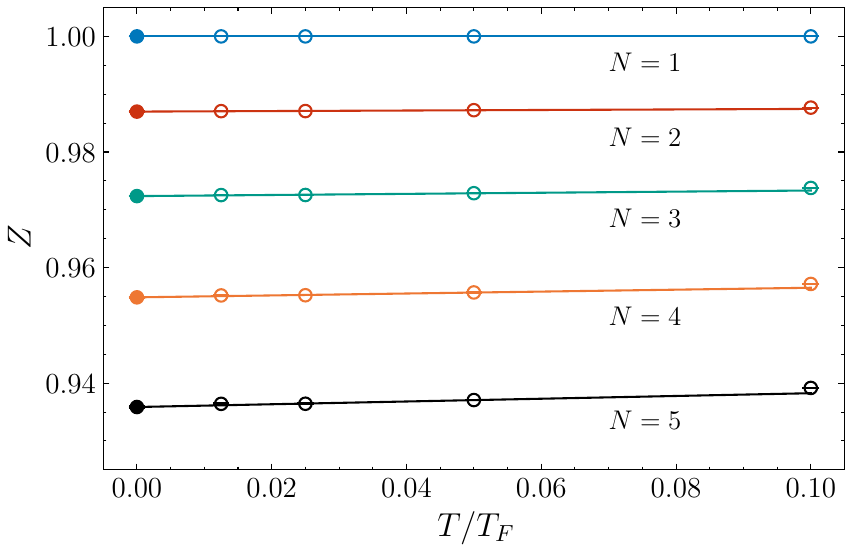}
    \caption{\label{fig:Tscale_Z}
    Temperature dependence of the quasiparticle residue $Z$ at $\rs=5$ ($\lambda_R=1.375$) for several renormalized orders.
    $Z$ is extracted from the frequency derivative of the imaginary part of the self-energy at the Fermi surface, $Z = [1 - \partial_\omega \mathrm{Im}\, \Sig(k,i\omega)|_{\kF,0}]^{-1}$.
    The linear-$T$ dependence at low temperature is consistent with Fermi-liquid theory.
    }
\end{figure}

\begin{figure}[t]
    \centering
    \includegraphics[width=0.95\columnwidth]{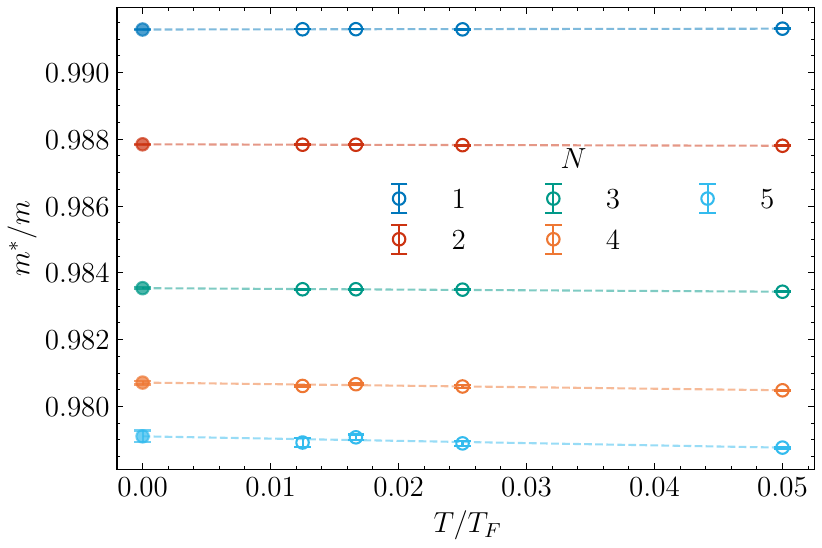}
    \caption{\label{fig:Tscale_vertex}
        Temperature dependence of $m^*/m$ from the vertex route at $\rs=5$ ($\lambda_R=1.375$) for several renormalized orders.
        The forward-scattering vertex $\Gamm_4$ is evaluated at the two lowest fermionic Matsubara frequencies ($\omega_0 = i\pi T$ and $\omega_{-1} = -i\pi T$) directly at $\mathbf{q}=0$, and the effective mass is extracted via Eq.~(3) of the main text.
        The linear-$T$ behavior mirrors that of the self-energy route and enables consistent $T\to 0$ extrapolation.
    }
\end{figure}

The Fermi-liquid parameters entering the effective-mass determinations are defined at zero temperature, but our diagrammatic calculations are performed in Matsubara formalism at finite temperature and extrapolated to $T=0$.
Figures~\ref{fig:Tscale_sigma}, \ref{fig:Tscale_Z}, and~\ref{fig:Tscale_vertex} demonstrate the low-temperature behavior of the key quantities at $\rs=5$ with $\lambda_R = 1.375$.

In the self-energy route, we compute the self-energy $\Sig(k, i\omega_n)$ on the Matsubara frequency axis at several low temperatures in the range $T/\TF \in [1/80, 1/20]$.
The quasiparticle residue $Z$ is extracted from the frequency derivative $\partial_\omega \mathrm{Im}\, \Sig(k,i\omega)|_{\kF,0}$ via a finite-difference approximation using the lowest two Matsubara frequencies.
The momentum derivative $\partial_k \Ree\, \Sig(k,\omega)|_{\kF,0}$ is obtained directly by Taylor-mode automatic differentiation of the diagrammatic computational graph with respect to the external momentum at $\kF$.
Both derivatives exhibit linear-$T$ dependence at low temperature, as expected in a Fermi liquid, and are extrapolated to $T=0$ by linear fits.
This linear dependence originates from evaluating the estimators at the lowest Matsubara frequencies $\pm i\pi T$, where quasiparticle damping contributes a term linear in $T$. The intrinsic finite-temperature corrections to three-dimensional (3D) Fermi-liquid parameters are nonanalytic but enter only at higher order [e.g., $\gamma(T)-\gamma(0)\propto T^{2}\ln T$ for the specific-heat coefficient~\cite{ChubukovMaslov2003,ChubukovMaslovMillis2006}] and are negligible over the fitted temperature window.

In the vertex route, the proper four-point vertex $\Gamm_{\sigma\sigma'}(\mathbf{k}_1\omega_{-1}, \mathbf{k}_2\omega_0; \mathbf{k}_1+\mathbf{q}\,\omega_0)$ is evaluated with all external momenta on the Fermi surface ($|\mathbf{k}_1| = |\mathbf{k}_2| = \kF$) at the two lowest fermionic Matsubara frequencies, and the forward-scattering limit is evaluated directly at $\mathbf{q}=0$, the $\omega$-limit being enforced by the distinct external Matsubara frequencies.
(The proper vertex excludes diagrams reducible by cutting a single interaction line. The direct $V_q$ contribution subtracted in Eq.~(2) of the main text is therefore absent, and the $\mathbf{q}=0$ evaluation is regular.)
The resulting Landau quasiparticle interaction $f_{\sigma\sigma'}(\theta)$ is projected onto its $l=1$ spin-symmetric component to obtain $F_1^s$, from which $m^*/m = 1 + F_1^s$ follows.
The temperature dependence of the vertex-route effective mass is likewise linear at low $T$ and is extrapolated to $T=0$ by the same procedure.

The observed linear-$T$ behavior across all perturbation orders supports the extrapolation protocol over the fitted temperature range.
The thermal correction to $m^*/m$ at $T/\TF = 1/40$ is approximately $\Delta(m^*/m) \sim 0.0002$, roughly an order of magnitude smaller than the total quoted uncertainty of ${\sim}0.002$ at $\rs=5$.
This small correction supports the use of finite-temperature calculations and permits comparison with ground-state properties at the quoted precision.

The systematic uncertainty associated with the $T\to 0$ extrapolation is estimated from the residuals of the linear fits and the extrapolation distance.
At the temperatures studied, the fitted linear behavior is well established, and the extrapolation uncertainty is subdominant compared to the truncation error and Monte Carlo statistical uncertainty.

\section{Density dependence and robustness of the auxiliary screening parameter}

\begin{figure*}[t]
    \centering
    \includegraphics[width=0.7\textwidth]{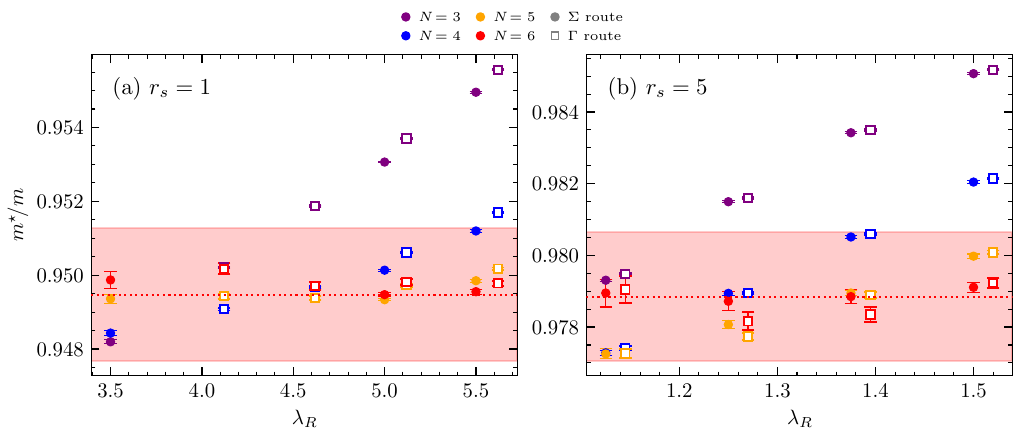}
    \caption{\label{fig:lam_scan}
        Robustness of $m^*/m$ with respect to the auxiliary Yukawa screening parameter $\lambda_R$ at $\rs=1$ and $\rs=5$.
        Filled symbols denote the self-energy route; open symbols denote the vertex route.
        The shaded bands indicate the final estimates and their uncertainties at each density.
        At sixth renormalized order, both routes exhibit weak residual $\lambda_R$ dependence over a broad range, with variations of less than $1\%$ in the plateau region.
        This confirms that the Yukawa screening is an auxiliary reorganization device rather than uncontrolled physical input, and that the extracted result is not contingent on a special choice of $\lambda_R$.
    }
\end{figure*}

Figure~\ref{fig:lam_scan} demonstrates the insensitivity of the converged effective-mass ratio to the choice of the auxiliary screening parameter $\lambda_R$ at two representative densities, $\rs=1$ (high density) and $\rs=5$ (intermediate density).

The Yukawa interaction $V_q(\lambda_R) = 4\pi e^2/(q^2 + \lambda_R)$ and its interaction counterterm reorganize the bare Coulomb series. The Coulomb interaction is recovered, and the result becomes independent of $\lambda_R$, when the series is summed to all orders.
At finite order, a residual $\lambda_R$ dependence remains. Its magnitude measures the residual finite-order sensitivity and the effectiveness of the reorganization.
The natural comparison is the Thomas--Fermi scale $q_{\rm TF}^2=4\pi e^2N_F$, where $N_F=mk_F/\pi^2$ is the total bare density of states. Using $k_Fa_B=(9\pi/4)^{1/3}/\rs$, where $a_B$ is the Bohr radius, gives $q_{\rm TF}^2a_B^2=(4/\pi)(9\pi/4)^{1/3}\rs^{-1}\simeq2.44/\rs$. Across $1\le\rs\le6$, the scanned plateau windows (Table~\ref{tab:lamR}) span $\lambda_R/q_{\rm TF}^2\approx1.4$--$3.4$, so $\sqrt{\lambda_R}=O(q_{\rm TF})$ throughout. Thus the useful finite-order scale follows the density-dependent screening scale without selecting a unique optimum. Values well below this scale leave the long-wavelength Coulomb singularity weakly reorganized, whereas much larger values increase the interaction counterterms and degrade the order-by-order behavior.

\begin{table}[t]
    \caption{\label{tab:lamR}
    Auxiliary-screening-parameter windows scanned at each density, in units of $a_B^{-2}$, and the corresponding ratio to the Thomas--Fermi scale $q_{\rm TF}^2=2.44/(\rs a_B^2)$.
    }
    \begin{ruledtabular}
        \begin{tabular}{c c c}
            $\rs$ & $\lambda_R$ scanned & $\lambda_R/q_{\rm TF}^2$ \\
            \hline
            1     & 3.5--8.0            & 1.4--3.3                 \\
            2     & 2.0--3.5            & 1.6--2.9                 \\
            3     & 1.5--2.25           & 1.8--2.8                 \\
            4     & 1.25--1.75          & 2.0--2.9                 \\
            5     & 1.125--1.5          & 2.3--3.1                 \\
            6     & 0.875--1.375        & 2.1--3.4                 \\
        \end{tabular}
    \end{ruledtabular}
\end{table}

At both densities shown, the sixth-order results from both routes vary by less than $1\%$ over a broad plateau of $\lambda_R$ values.
The two routes remain mutually consistent across the entire scanned range, providing a nontrivial cross-check of the diagrammatic content.
The final values use representative $\lambda_R$ values within these broad plateaus, including $1.375$ at $\rs=5$, chosen for stable order-by-order convergence.

The residual $\lambda_R$ sensitivity at sixth order is incorporated into the uncertainty budget by taking the maximum variation of $m^*/m$ over the plateau region as a systematic-error contribution.
This contribution is subdominant compared to the truncation error estimated from the order-by-order analysis, confirming that the series is well controlled at the orders reached.

\section{Uncertainty budget}

The quoted uncertainties in the main text combine four sources of systematic and statistical error, added in quadrature:

\textbf{(1) Monte Carlo statistical errors.}
Each diagram is evaluated by Monte Carlo (MC) integration over the internal loop variables using \texttt{MCIntegration.jl}.
The statistical uncertainty is estimated from the sample variance of the Monte Carlo iterations and is typically in the range $\Delta(m^*/m)_{\mathrm{MC}} \sim 0.0001$--$0.0003$ at sixth order, depending on the density and route.

\textbf{(2) Truncation error.}
The systematic uncertainty associated with truncating the renormalized series at sixth order is estimated from the variation across the last three perturbation orders ($N=4, 5, 6$), as described in Sec.~I.
This is the dominant contribution to the total uncertainty at most densities, typically $\Delta(m^*/m)_{\mathrm{trunc}} \sim 0.001$--$0.002$.

\textbf{(3) Low-temperature extrapolation.}
The systematic uncertainty in the $T\to 0$ extrapolation is estimated from the residuals of the linear fits to the low-$T$ data and the extrapolation distance.
At the temperatures studied ($T/\TF \lesssim 1/20$), the linear Fermi-liquid regime is well established, and this contribution is subdominant: $\Delta(m^*/m)_{T} \sim 0.0001$--$0.0003$.

\textbf{(4) Residual $\lambda_R$ sensitivity.}
The systematic uncertainty associated with the choice of the auxiliary screening parameter is estimated from the maximum variation of $m^*/m$ over the plateau region in the $\lambda_R$ scans (Fig.~\ref{fig:lam_scan}).
At sixth order, this contribution is likewise subdominant: $\Delta(m^*/m)_{\lambda} \sim 0.0002$--$0.0005$.

The combined values plotted in Fig.~1 of the main text are the inverse-variance weighted means of the two routes. Because contributions (2)--(4) are estimated from the same renormalized series, the same $\lambda_R$ scans, and the same $T\to0$ protocol for both routes, they are treated as fully correlated in the combination. Each plotted uncertainty is conservatively assigned as the larger route uncertainty rather than reduced by averaging.
The four contributions are combined in quadrature to obtain the total quoted uncertainty:
\begin{equation}
    \Delta(m^*/m)_{\mathrm{total}} = \sqrt{\Delta_{\mathrm{MC}}^2 + \Delta_{\mathrm{trunc}}^2 + \Delta_T^2 + \Delta_\lambda^2} \,.
\end{equation}
At $\rs=5$, for example, the typical breakdown is $\Delta_{\mathrm{MC}} \sim 0.0002$, $\Delta_{\mathrm{trunc}} \sim 0.0015$, $\Delta_T \sim 0.0002$, $\Delta_\lambda \sim 0.0003$, yielding $\Delta_{\mathrm{total}} \sim 0.0015$--$0.0020$ depending on the route.
The range reflects the slightly different error contributions in the two routes.
The truncation contribution dominates because measurable order-to-order variation remains through sixth order and is conservatively used to bound residual higher-order effects.
The mutual agreement of the two routes within these uncertainties provides a nontrivial internal consistency check and supports the quoted uncertainty estimates.

\section{Systematic sensitivities of finite-cell excitation-energy estimates of \texorpdfstring{$m^*$}{m*}}\label{sec:dmc}

{
We examine systematic sensitivities of the finite-cell excitation-energy procedure used in the DMC study of Ref.~\cite{azadi2021}, using the published results and analytic long-wavelength finite-size theory~\cite{holzmann2011finite}.
The central methodological question is whether the Fermi-surface slope obtained from a global fit to finite-cell addition and removal energies converges to that of the thermodynamic-limit quasiparticle dispersion.
We then assess finite-size, trial-state, and algorithmic sensitivities of this extraction.
This analysis does not assign a corrected DMC mass or attribute the discrepancy to a unique source.

Reference~\cite{azadi2021} constructs an energy band from fixed-node total-energy differences, $\varepsilon(k)=E^{+}(k)-E_0$ for $k>\kF$ and $\varepsilon(k)=E_0-E^{-}(k)$ for $k<\kF$, where $E^{\pm}(k)$ are $(N{\pm}1)$-electron energies with a plane-wave orbital added or removed.
The calculation uses closed-shell $\Gamma$-point cells ($N=54$--$274$), up to twenty momenta over $0\le k\lesssim2\kF$, a global Pad\'e fit whose slope at $\kF$ yields $m^*$, and an extrapolation of $m^*(N)$ linear in $1/N$ over $N=130$--$274$ (Fig.~3 of Ref.~\cite{azadi2021}).
Because the band values nearest $\kF$ remain irregular in finite cells---they inherit the Hartree--Fock logarithmic anomaly, as noted in Ref.~\cite{azadi2021}---the global fit necessarily uses excitations away from the Fermi surface.
The extracted slope is therefore the derivative of an interpolation constrained over a broad momentum range rather than a directly resolved local Fermi-surface derivative.
Reference~\cite{azadi2021} found similar results from Pad\'e and quartic fits, which establishes stability against these two fitting forms but does not test convergence under a momentum window that shrinks toward $\kF$ as $N$ increases.
The following three sensitivities are therefore not tested by agreement between the two fit forms.

\emph{(i) Finite-cell excitation energies and the quasiparticle dispersion.}
The quasiparticle dispersion is defined by the pole or resonance of the thermodynamic-limit one-particle Green's function near $\kF$~\cite{holzmann2023static}.
At finite $N$, the addition and removal energies are fixed-node estimates of discrete excitations in the $(N{\pm}1)$-particle sectors.
Because the quasiparticle width grows as $\Gamma(k)\propto(k-\kF)^2$~\cite{GiulianiVignale2005}, the accuracy and convergence rate of this correspondence away from $\kF$ have not been established at the precision relevant here.
Consequently, the broad-window derivative has not been shown to reproduce the joint thermodynamic and Fermi-surface limit at that precision.
Within the Slater--Jastrow variational Monte Carlo (VMC) calculations of Ref.~\cite{holzmann2023static}, the excitation-band estimates decrease from $m_{\mathrm{eff}}/m\simeq0.93$ to $0.83$ over $\rs=1$--$10$, whereas the static Green's-function estimates increase from approximately $0.96$ to $1.13$; the corresponding backflow static estimates increase from approximately $1.00$ to $1.28$.
Those authors attributed the qualitative difference mainly to the extraction methodology rather than to DMC projection.
A companion two-dimensional (2D) DMC study likewise states that the mapping between finite-cell addition/removal energies away from $\kF$ and quasiparticle excitations is not guaranteed and remains an open question~\cite{azadi2025qmc2d}.

\emph{(ii) Long-wavelength Coulomb finite-size contribution.}
For a Fermi-surface quasiparticle mass defined through derivatives of the one-particle self-energy, Ref.~\cite{holzmann2011finite} derives a leading long-wavelength correction for this estimator:
\begin{equation}\label{eq:deltaFS}
    \Delta_N=-c_3\left(\frac{\rs}{3}\right)^{1/2}
    \left(\frac{3}{4\pi N}\right)^{1/3},
    \qquad c_3\simeq1.221,
\end{equation}
which enters the finite-size relation
\begin{equation}\label{eq:meffFS}
    \left(\frac{m}{m^*}-1\right)_\infty
    \simeq
    \left(\frac{m}{m^*}-1\right)_N(1+\Delta_N Z).
\end{equation}
Here $N^{-1/3}\propto L^{-1}$ at fixed density, reflecting the missing long-wavelength plasmon sector.
The derivation uses grand-canonical twist averaging to remove one-body shell effects and isolates the long-wavelength quadrature error in the Fermi-surface self-energy derivatives.
Reference~\cite{azadi2021}, by contrast, extrapolates the slopes of global $\Gamma$-point finite-cell excitation-band fits linearly in $1/N$.
Because the estimators and boundary-condition treatments differ, Eq.~\eqref{eq:meffFS} does not establish the asymptotic scaling law for those fitted masses and cannot be applied as a direct correction.
It identifies a slowly decaying long-wavelength contribution for the self-energy estimator, but the cited analysis does not establish whether an analogous term is present in the broad-band estimator over the simulated sizes.
A companion 2D study reports sensitivity to the extrapolation exponent and to how the excitation band is constructed~\cite{azadi2025qmc2d}, but these observations do not determine the finite-size scaling in three dimensions.

\emph{(iii) Trial-state and algorithmic sensitivity.}
Reference~\cite{azadi2021} compared Slater--Jastrow and Slater--Jastrow--backflow masses only at $N=178$ and two densities, finding shifts of $+0.006$ at $\rs=1$ and $+0.03$ at $\rs=10$.
These small shifts indicate modest trial-state sensitivity at the tested 3D points, but a single size and two densities do not bound its effect on the thermodynamic extrapolation.
In the companion 2D calculation, backflow changed both the magnitude and density trend, and the extracted mass was also sensitive to the DMC time step~\cite{azadi2025qmc2d}.
Although these 2D shifts cannot be transferred quantitatively to 3D, they show that a generic cancellation of fixed-node and algorithmic errors between ground- and excited-state energies cannot be assumed for this estimator.

Taken together, these analyses identify three systematic sensitivities not included in the reported statistical error bars: the relation between a global finite-cell excitation band and the thermodynamic quasiparticle dispersion, the estimator dependence of the finite-size extrapolation, and incompletely quantified trial-state and algorithmic effects.
They can affect an effective-mass determination at the precision relevant to the discrepancy, but they neither provide a unique correction nor establish its sole origin.
These considerations also do not resolve the smaller spread among Green's-function-based determinations: at $\rs=5$, for example, the backflow-VMC value $1.09(3)$ of Ref.~\cite{holzmann2023static} lies above the present $0.979(2)$, and resolving this difference remains an open task.
}
\bibliographystyle{apsrev4-2}
\bibliography{references}

%% file: fig_diagrams.tex
% Requires: \usepackage{tikz}
% \usetikzlibrary{decorations.pathmorphing,decorations.markings,arrows.meta,calc,shapes.misc}

\begin{tikzpicture}[
        fermion/.style={thick,
                postaction={decorate,
                        decoration={markings,
                                mark=at position 0.55 with
                                    {\arrow{Stealth[length=4.5pt,width=3.5pt]}}}}},
        yukawa/.style={thick,
                decorate,decoration={snake,amplitude=1.1pt,segment length=4.5pt,
                        pre length=2.5pt,post length=2.5pt}},
        ctmu/.style={draw,thick,circle,fill=white,inner sep=0pt,minimum size=5pt,
                path picture={\draw[thick]
                        (path picture bounding box.north west)--(path picture bounding box.south east)
                        (path picture bounding box.south west)--(path picture bounding box.north east);}},
        ctV/.style={fill=black,rectangle,inner sep=0pt,minimum size=3.5pt},
        dot/.style={circle,fill=black,inner sep=0pt,minimum size=2.5pt},
        lbl/.style={font=\scriptsize},
        olbl/.style={font=\scriptsize,text=black!60},
        every node/.style={font=\normalsize},
        % 【修改点1】：全局比例放大，从 0.55cm 提升至 0.65cm，使图整体更饱满
        x=0.65cm, y=0.65cm
    ]

    % ==========================================================
    %  (a)  SELF-ENERGY  Σ  (Row 1 & Row 2)
    % ==========================================================
    \node[font=\small\bfseries,anchor=east] at (0,1.6) {(a)};

    %%% Row 1 Labels & Grid (位置更加紧凑)
    \node[olbl] at (2.0, 1.4) {order 1};
    \node[olbl] at (8.0, 1.4) {order 2 CT};
    \draw[black!50,densely dashed] (4.0,-0.4) -- (4.0,1.5);

    %%% --- Diagram 1: Fock ---
    \begin{scope}[shift={(1.0,0)}]
        \coordinate (a) at (0,0);
        \coordinate (b) at (2.0,0);
        \draw[fermion] (-0.6,0) -- (a);
        \draw[fermion] (a) -- (b);
        \draw[fermion] (b) -- (2.6,0);
        \draw[yukawa] (a) to[bend left=80] (b);
        \node[dot] at (a) {}; \node[dot] at (b) {};
    \end{scope}

    %%% --- Diagram 2: CT Fock + δg ---
    \begin{scope}[shift={(5.0,0)}]
        \coordinate (a) at (0,0);
        \coordinate (b) at (2.0,0);
        \coordinate (m) at (1.0,0);
        \draw[fermion] (-0.6,0) -- (a);
        \draw[fermion] (a) -- (m);
        \draw[fermion] (m) -- (b);
        \draw[fermion] (b) -- (2.6,0);
        \draw[yukawa] (a) to[bend left=80] (b);
        \node[ctmu] at (m) {};
        \node[lbl,below=3pt] at (m) {$\delta \mu^{(1)}$};
        \node[dot] at (a) {}; \node[dot] at (b) {};
    \end{scope}

    %%% --- Diagram 3: CT Fock + δV ---
    \begin{scope}[shift={(9.0,0)}]
        \coordinate (a) at (0,0);
        \coordinate (b) at (2.0,0);
        \draw[fermion] (-0.6,0) -- (a);
        \draw[fermion] (a) -- (b);
        \draw[fermion] (b) -- (2.6,0);
        \draw[yukawa] (a) to[bend left=80] node[pos=0.5, ctV] (ym) {} (b);
        % \node[lbl,above=3pt] at (ym) {$\delta_V^{(1)}$};
        % \node[lbl,above=3pt] at (ym) {$\lambda_R/4\pi e^2$};
        \node[dot] at (a) {}; \node[dot] at (b) {};
    \end{scope}

    %%% Row 2 Labels
    \node[olbl] at (6.0, -1.5) {order 2};

    %%% --- Diagram 4: Bubble self-energy ---
    % 【修改点2】：Y轴间距大幅压缩 (shift y= -3.0)
    \begin{scope}[shift={(1.0,-3.0)}]
        \coordinate (v1) at (0,0);
        \coordinate (v2) at (2.0,0);
        % \coordinate (v3) at (0.5,1.2);
        % \coordinate (v4) at (1.5,1.2);
        \coordinate (v3) at (0.5,1.0);
        \coordinate (v4) at (1.5,1.0);
        \draw[fermion] (-0.6,0) -- (v1);
        \draw[fermion] (v1) -- (v2);
        \draw[fermion] (v2) -- (2.6,0);
        \draw[yukawa] (v1) -- (v3);
        \draw[yukawa] (v2) -- (v4);
        \draw[fermion] (v3) to[bend left=60] (v4);
        \draw[fermion] (v4) to[bend left=60] (v3);
        \node[dot] at (v1) {}; \node[dot] at (v2) {};
        \node[dot] at (v3) {}; \node[dot] at (v4) {};
    \end{scope}

    %%% --- Diagram 5: Nested / rainbow ---
    \begin{scope}[shift={(5.0,-3.0)}]
        \coordinate (v1) at (0,0);
        \coordinate (v2) at (0.6,0);
        \coordinate (v3) at (1.4,0);
        \coordinate (v4) at (2.0,0);
        \draw[fermion] (-0.6,0) -- (v1);
        \draw[fermion] (v1) -- (v2);
        \draw[fermion] (v2) -- (v3);
        \draw[fermion] (v3) -- (v4);
        \draw[fermion] (v4) -- (2.6,0);
        \draw[yukawa] (v1) to[bend left=80] (v4);
        \draw[yukawa] (v2) to[bend left=80] (v3);
        \node[dot] at (v1) {}; \node[dot] at (v2) {};
        \node[dot] at (v3) {}; \node[dot] at (v4) {};
    \end{scope}

    %%% --- Diagram 6: Crossed ---
    \begin{scope}[shift={(9.0,-3.0)}]
        \coordinate (v1) at (0,0);
        \coordinate (v2) at (0.6,0);
        \coordinate (v3) at (1.4,0);
        \coordinate (v4) at (2.0,0);
        \draw[fermion] (-0.6,0) -- (v1);
        \draw[fermion] (v1) -- (v2);
        \draw[fermion] (v2) -- (v3);
        \draw[fermion] (v3) -- (v4);
        \draw[fermion] (v4) -- (2.6,0);
        \draw[yukawa] (v1) to[bend left=75] (v3);
        \draw[yukawa] (v2) to[bend right=75] (v4);
        \node[dot] at (v1) {}; \node[dot] at (v2) {};
        \node[dot] at (v3) {}; \node[dot] at (v4) {};
    \end{scope}

    % ==========================================================
    %  (b)  PROPER FOUR-POINT VERTEX  Γ₄  (Row 3 & Row 4)
    % ==========================================================
    % 【修改点3】：去掉了 (a)(b) 之间的虚线，进一步拉近距离
    \node[font=\small\bfseries,anchor=east] at (0,-5.0) {(b)};

    %%% Row 3 Labels & Grid 
    \node[olbl] at (4.0, -5.0) {order 1};
    \node[olbl] at (10.0,-5.0) {order 2 CT};
    \draw[black!50,densely dashed] (8.0,-7.6) -- (8.0,-5.0);

    %%% --- Diagram 1: Direct ---
    \begin{scope}[shift={(2.0,-6.6)}]
        \coordinate (vl) at (-0.8,0);
        \coordinate (vr) at (0.8,0);
        \draw[fermion] (-1.4,-0.8) -- (vl);
        \draw[fermion] (vl) -- (-1.4, 0.8);
        \draw[fermion] ( 1.4,-0.8) -- (vr);
        \draw[fermion] (vr) -- ( 1.4, 0.8);
        \draw[yukawa] (vl) -- (vr);
        \node[dot] at (vl) {}; \node[dot] at (vr) {};
    \end{scope}

    %%% --- Diagram 2: Exchange (带弧度的出射外腿) ---
    \begin{scope}[shift={(6.0,-6.6)}]
        \coordinate (vl) at (-0.8,0);
        \coordinate (vr) at (0.8,0);
        \draw[fermion] (-1.4,-0.8) -- (vl);
        \draw[fermion] (1.4,-0.8) -- (vr);
        % 【修改点4】：出射腿采用平滑的贝塞尔曲线交叉
        % \draw[fermion] (vl) to[out=45, in=210] (1.4, 0.8);
        % \draw[fermion] (vr) to[out=135, in=-30] (-1.4, 0.8);
        \draw[fermion] (vl) to[out=45, in=190] (1.4, 0.8);
        \draw[fermion] (vr) to[out=135, in=-10] (-1.4, 0.8);
        \draw[yukawa] (vl) -- (vr);
        \node[dot] at (vl) {}; \node[dot] at (vr) {};
    \end{scope}

    %%% --- Diagram 3: Counterterm on Direct ---
    \begin{scope}[shift={(10.0,-6.6)}]
        \coordinate (vl) at (-0.8,0);
        \coordinate (vr) at (0.8,0);
        \draw[fermion] (-1.4,-0.8) -- (vl);
        \draw[fermion] (vl) -- (-1.4, 0.8);
        \draw[fermion] ( 1.4,-0.8) -- (vr);
        \draw[fermion] (vr) -- ( 1.4, 0.8);
        \draw[yukawa] (vl) -- node[pos=0.5, ctV] (vm) {} (vr);
        % \node[lbl,above=3pt] at (vm) {$\delta_V^{(1)}$};
        \node[dot] at (vl) {}; \node[dot] at (vr) {};
    \end{scope}

    %%% Row 4 Labels
    % \node[olbl] at (6.0, -8.7) {order 2};
    \node[olbl] at (6.0, -8.2) {order 2};

    %%% --- Diagram 4: Bubble ---
    \begin{scope}[shift={(2.0,-9.8)}]
        \coordinate (vl) at (-1.1,0);
        \coordinate (vr) at (1.1,0);
        \coordinate (vml) at (-0.5,0);
        \coordinate (vmr) at (0.5,0);
        \draw[fermion] (-1.6,-0.8) -- (vl);
        \draw[fermion] (vl) -- (-1.6, 0.8);
        \draw[fermion] ( 1.6,-0.8) -- (vr);
        \draw[fermion] (vr) -- ( 1.6, 0.8);
        \draw[yukawa] (vl) -- (vml);
        \draw[yukawa] (vmr) -- (vr);
        \draw[fermion] (vml) to[bend left=70] (vmr);
        \draw[fermion] (vmr) to[bend left=70] (vml);
        \node[dot] at (vl) {}; \node[dot] at (vr) {};
        \node[dot] at (vml) {}; \node[dot] at (vmr) {};
    \end{scope}

    %%% --- Diagram 5: Exchange Vertex Correction (提前，并加入外腿弧度) ---
    % 【修改点5】：与Box图互换了位置
    \begin{scope}[shift={(6.0,-9.8)}]
        \coordinate (A) at (-0.8, -0.6);
        \coordinate (C) at (-0.8,  0.6);
        \coordinate (B) at (-0.3,  0.0);
        \coordinate (D) at ( 0.6,  0.0);

        \draw[fermion] (-1.4, -0.9) -- (A);
        \draw[fermion] ( 1.4, -0.9) -- (D);

        % 平滑的出射交叉腿
        % \draw[fermion] (C) to[out=15, in=210] (1.4, 0.9);
        % \draw[fermion] (D) to[out=135, in=-30] (-1.4, 0.9);
        \draw[fermion] (C) to[out=15, in=190] (1.4, 0.9);
        \draw[fermion] (D) to[out=135, in=-10] (-1.4, 0.9);

        % 左侧内部顶角修正
        \draw[fermion] (A) -- (B);
        \draw[fermion] (B) -- (C);
        \draw[yukawa]  (A) to[bend left=55] (C);
        \draw[yukawa] (B) -- (D);

        \node[dot] at (A) {}; \node[dot] at (B) {};
        \node[dot] at (C) {}; \node[dot] at (D) {};
    \end{scope}

    %%% --- Diagram 6: Box (靠后放置) ---
    \begin{scope}[shift={(10.0,-9.8)}]
        \coordinate (bl) at (-0.8,-0.6);
        \coordinate (br) at (0.8,-0.6);
        \coordinate (tl) at (-0.8,0.6);
        \coordinate (tr) at (0.8,0.6);
        \draw[fermion] (-1.4,-0.9) -- (bl);
        \draw[fermion] (tl) -- (-1.4, 0.9);
        \draw[fermion] ( 1.4,-0.9) -- (br);
        \draw[fermion] (tr) -- ( 1.4, 0.9);
        \draw[fermion] (bl) -- (tl);
        \draw[fermion] (br) -- (tr);
        \draw[yukawa] (bl) -- (br);
        \draw[yukawa] (tl) -- (tr);
        \node[dot] at (bl) {}; \node[dot] at (br) {};
        \node[dot] at (tl) {}; \node[dot] at (tr) {};
    \end{scope}

    % ==========================================================
    %  LEGEND
    % ==========================================================
    % 图例也相应上移
    \begin{scope}[shift={(0.5,-12.0)}]
        \draw[fermion] (0,0) -- (1.0,0);
        \node[anchor=west,lbl] at (1.1,0) {$G_0(\mu_R)$};

        \draw[yukawa]  (3.8,0) -- (4.8,0);
        \node[anchor=west,lbl] at (4.9,0) {$V_q(\lambda_R)$};

        \node[ctmu] at (7.8,0) {};
        \node[anchor=west,lbl] at (8.1,0) {$\delta\mu$};

        \node[ctV]  at (10.0,0) {};
        \node[anchor=west,lbl] at (10.2,0) {$\frac{\lambda_R}{4\pi e^2}$};
    \end{scope}

\end{tikzpicture}